\documentclass[journal]{IEEEtran}
\usepackage[colorinlistoftodos]{todonotes} 
\usepackage{todonotes}
\usepackage{graphicx}
\usepackage{amsmath}
\usepackage{url}
\usepackage{cite}
\usepackage{xcolor}
\usepackage{siunitx}
\usepackage{tabularx}
\usepackage[caption=false,font=footnotesize]{subfig}
\usepackage{adjustbox}
\usepackage{booktabs} 
\usepackage{multirow}
\usepackage{lipsum}  
\begin{document}

\title{Stochastic Capacity Accreditation: Incentivizing Resource
Adequacy under Weather Uncertainty 
}

\author{ Hangrui~Liu,~\IEEEmembership{Student~Member,~IEEE}, Shen~Wang,~\IEEEmembership{Member,~IEEE}, Audun~Botterud,~\IEEEmembership{Senior~Member,~IEEE}, and~Miguel~F.~Anjos,~\IEEEmembership{Senior~Member,~IEEE} \thanks{Hangrui~Liu and Miguel~F.~Anjos are with  the School of Mathematics and Maxwell Institute for Mathematical Sciences, University of Edinburgh, Edinburgh, EH9 3FD, United Kingdom.} \thanks{Hangrui~Liu is also with the MIT Energy Initiative, Massachusetts Institute of Technology, Cambridge, MA, USA.} 
\thanks{Miguel F. Anjos is also with GERAD HEC Montr\'eal, 3000, chemin de la C\^ote-Sainte-Catherine, Montr\'eal (Qu\'ebec) Canada H3T 2A7.}
\thanks{Shen~Wang is with the Center for Energy and Environmental Policy Research, Massachusetts Institute of Technology, Cambridge, MA, USA, and the MIT Energy Initiative, Massachusetts Institute of Technology, Cambridge, MA, USA.} \thanks{Audun~Botterud is with the Laboratory for Information and Decision Systems, Massachusetts Institute of Technology, Cambridge, MA, USA.} }

\maketitle

\begin{abstract}
High penetrations of variable renewable energy introduce significant resource adequacy challenges, particularly when weather-driven uncertainty affects renewable availability, electricity demand, and the effective capacity of thermal generators. Existing capacity credit accreditation methods often neglect these correlated weather effects, which may overstate firm capacity, distort long-term investment decisions, and weaken reliability outcomes in electricity market with price caps. This paper proposes a two-stage stochastic optimization framework for capacity accreditation that explicitly captures uncertainty in wind, solar, and temperature-dependent thermal derating. Using five years of ERCOT demand and renewable availability data, we compare the proposed stochastic capacity credit method with deterministic and average accreditation approaches, and quantify the impact of alternative accreditation methods on reliability, investment incentives, and the value of weather information. The results show that incorporating weather uncertainty yields more informative capacity credits and more reliable investment signals. In contrast, deterministic and averaged approaches can distort resource expansion decisions and produce materially worse reliability outcomes. These findings demonstrate the importance of explicitly accounting for weather uncertainty in capacity accreditation and long-term resource adequacy planning.

\begin{IEEEkeywords}
Power systems planning, resource adequacy,  capacity accreditation, weather uncertainty
\end{IEEEkeywords}

\end{abstract}

\section*{Nomenclature} 

\subsection{Sets}
\noindent
\begin{tabular}{@{}l p{0.65\columnwidth}@{}}
$T$ 
& Set of time periods, $T=\{1,2,\ldots,N_T\}$. \\

$G^F,\,G^W,\,G^{PV},\,G$ 
& Sets of fossil-fueled conventional, wind, solar, and all generation units/types, with $G = G^F \cup G^W \cup G^{PV}$. \\

$S$ 
& Set of all storage units/types, indexed by $s \in S$. \\

$\Omega$ 
& Set of scenarios, indexed by $\omega \in \Omega$. \\
\end{tabular}

\subsection{Parameters}
\noindent
\begin{tabular}{@{}l p{0.68\columnwidth}@{}}
$\alpha_s$ 
& Initial state-of-charge coefficient for storage $s \in S$ (unitless). \\

$CC_g,\,CC_s$ 
& Capacity credits of generation resource $g \in G$, storage resource $s \in S$ (unitless). \\

$D_{t,w}$    & System demand at time $t \in T$ for scenario $ \omega$ [MW]. \\

$HD$          & Energy storage duration (hours). \\

$PD$          & Peak demand [MW]. \\

$C_g,\,C_s$ 
& Investment costs of generator $g \in G$ and storage resource $s \in S$, respectively [\$/MW]. \\

$VC_g,\,VSC_s$ 
& Variable costs of generator $g \in G$, storage resource $s \in S$ [\$/MWh]. \\

$PRM$          & Minimum planning reserve margin requirement (unitless or \%). \\

$SC_s,\,SD_s$ 
& Maximum charging and discharging rate coefficients for storage $s \in S$ (unitless). \\

$TDF_{g,t,\omega}$ 
& Temperature-dependent derating factor of conventional generator $g \in G^F$ at time $t \in T$ in scenario $\omega \in \Omega$ (unitless). \\

$VOLL$        & Value of lost load [\$/MWh]. \\

$\phi_{g,t,\omega}$
& Availability factor of wind or solar generation resource $g \in G^W \cup G^{PV}$ at time $t \in T$ in scenario $\omega \in \Omega$ (unitless). \\

$\eta_c,\,\eta_d$ 
& Charging and discharging efficiencies of storage (unitless). \\

$\pi_{\omega}$ 
& Probability of scenario $\omega \in \Omega$. \\
\end{tabular}

\subsection{Decision Variables}
\noindent
\begin{tabularx}{\linewidth}{@{}l X@{}}
$c_{s,t,\omega}$ 
& Charging power from the grid to storage unit $s \in S$ at time $t \in T$ for scenario $\omega$ [MW]. \\

$d_{s,t,\omega}$ 
& Discharging power from storage unit $s \in S$ at time $t \in T$ for scenario $\omega$ [MW]. \\

$x_g,\,x_s$ 
& Installed capacity of generation resource $g \in G$ and storage resource $s \in S$, respectively [MW]. \\

$p_{g,t,\omega}$ 
& Energy output of generator $g \in G$ at time $t \in T$ for scenario $\omega$, with $p_{g,t,\omega} \geq 0$ [MWh]. \\

$ue_{t,\omega}$ 
& Unserved energy at time $t \in T$ for scenario $\omega$ [MWh]. \\

$e_{s,t,\omega}$ 
& State of charge (energy level) of storage unit $s \in S$ at the end of time $t$ for scenario $\omega$ [MWh]. \\
\end{tabularx}

\section{Introduction}

Power systems worldwide are undergoing a profound transformation driven by decarbonization and the rapid expansion of variable renewable energy (VRE) resources \cite{SEPULVEDA20182403,en14092408}. This transition is occurring in parallel with increasing climate-related weather variability. According to the Intergovernmental Panel on Climate Change (IPCC), average global temperatures are projected to continue rising relative to pre-industrial levels, with important implications for electricity demand, generator performance, and system reliability \cite{masson2019global}. Accordingly, long-term resource adequacy planning must increasingly account for the effects of weather on renewable output, load, and the effective availability of conventional generation.

Weather affects all major classes of power system resources. Wind generation depends directly on wind speed, which varies across time and location. Solar generation is similarly sensitive to meteorological conditions such as irradiance and cloud cover \cite{perez2019climate,crook2011climate,fenger2007impacts}. Conventional thermal generators are also weather-sensitive, because ambient conditions affect both equipment performance and outage behavior \cite{MURPHY2019113513}. These effects are often correlated with system stress: for example, extreme temperatures can simultaneously increase demand and reduce thermal availability, while renewable availability may rise or fall depending on the weather regime \cite{champaign_zba_086AT23_2023,sarmah2023comprehensive}. Such joint and correlated weather impacts create multidimensional resource adequacy risks that are not well represented by traditional deterministic planning approaches.

\subsection{Resource adequacy and capacity markets}
Resource adequacy is the ability of a power system to meet demand under a wide range of operating conditions \cite{denholm_resource_adequacy_2025}. In both centralized planning and wholesale electricity markets, achieving long-term adequacy requires converting reliability targets into economic signals that support efficient investment in capacity. For example, besides the traditional one-day-in-ten-years loss-of-load expectation (LOLE) criterion, the NERC 2025 Long-Term Reliability Assessment (LTRA) also uses an annual NEUE threshold of 20 ppm, equivalent to 0.002\% of annual energy demand, as a high-risk threshold for assessing energy adequacy \cite{nerc_ltra_2025}. In practice, however, energy-only markets may fail to provide sufficient scarcity revenues to sustain adequate long-term investment, particularly when price caps limit cost recovery during stressed conditions. This ``missing money'' problem motivates capacity mechanisms, including capacity markets, which provide explicit compensation for accredited capacity and thereby help support reliability \cite{cramton2013capacity}.

A central concept in both adequacy planning and capacity-market design is the \emph{capacity credit} (CC), which measures a resource's contribution to system reliability \cite{5589606,SSENGONZI2022100033}. In planning models, CC converts installed nameplate capacity into an equivalent firm contribution that can be compared against an adequacy requirement, often represented through a planning reserve margin. 

In capacity markets, CCs determine the share of installed capacity eligible for payments. CCs therefore directly affect both investment incentives and reliability outcomes: overestimation may understate adequacy risk, whereas underestimation may induce over-procurement and higher system cost. As weather variability increases, accreditation methods that ignore uncertainty may distort both future capacity payments and long-term planning decisions \cite{mantegna2025electric,sun2025navigating,ZUO2025108771}.

\subsection{Existing accreditation practices and their limitations}

Historically, capacity accreditation was developed in systems dominated by conventional generation. For thermal units, many system operators rely on deterministic or historical-performance-based metrics such as unforced capacity (UCAP), which derates installed capacity using historical forced outage rates.  Broader probabilistic resource adequacy assessments may use Monte Carlo simulation to evaluate system-level reliability metrics.

However, recent work shows that many conventional approaches still do not adequately capture temperature-dependent deratings and correlated outage behavior under extreme weather, for example, \cite{MURPHY2019113513} shows that adequacy models often treat generator failures as independent of ambient conditions and finds that temperature-dependent outages can materially increase required reserve margins. EPRI and ISO-NE similarly note that temperature-related deratings and ambient-condition have effects on thermal performance \cite{epri2022resource,potomac_isone_emm_2024}. Nevertheless, thermal availability is still frequently represented using stationary outage assumptions or simplified historical derating factors. This can bias adequacy assessments, since extreme temperatures may simultaneously raise outage risk, reduce thermal output, and coincide with high demand. These findings suggest that thermal capacity credits should also be evaluated under weather uncertainty, rather than treated as fixed or weather-invariant.

VRE output varies substantially across hours, seasons, and weather regimes, so its adequacy value cannot be captured adequately by outage-based metrics. A widely used probabilistic metric for VRE is Effective Load Carrying Capability (ELCC), originally proposed by Garver for traditional generation \cite{4073133}. ELCC measures the additional demand that can be served after adding a candidate resource while maintaining the same reliability level. Related measures such as equivalent firm capacity (EFC) are also used in practice \cite{Zachary_2022,4806129}. Many system operators now use ELCC-based methodologies for VRE accreditation; for example, SPP, PJM, MISO, and NESO have adopted ELCC-based approaches \cite{spp_elcc_study_scope_2025,spp_landmark_2025, isone_mri_memo_2025, pjm_elcc_education_2024,pjm_manual_20a_2025, NationalGridESO2023, miso_resource_accreditation_v21_2024}.

Despite this progress, important methodological limitations remain in how existing capacity accreditation approaches incorporate weather information. In most current methods, weather enters the calculation through historical realizations, predefined reliability-critical periods, or historical averages, rather than as an explicit source of uncertainty within the planning problem. In the remainder of this section, we summarize three CC approaches---\emph{Deterministic Capacity Credits} (DCC), \emph{Average Capacity Credits} (AVGCC), and \emph{Expected-Value Capacity Credits} (EVCC)---that differ in how weather information is incorporated into the accreditation calculation.

In this paper, DCC refers to capacity credits derived from fixed historical realizations. In practice, several ELCC-based studies and ISO/RTO implementations evaluate accreditation using deterministic historical weather years, often analyzing different deterministic periods separately and then combining the results across them
\cite{miso_resource_accreditation_v21_2024}, \cite{9473022}, \cite{9729557}. This represents an improvement over purely deterministic approximations, but it still ties accreditation outcomes to a finite set of historical realizations and may not fully capture the range of different weather conditions relevant to resource adequacy.

Moving beyond deterministic realizations, some approaches attempt to incorporate multiple weather scenarios. A related limitation is that many existing approaches incorporate weather information only through ex post analysis. That is, reliability contributions are evaluated conditional on different weather scenarios, and the resulting EUE values are then averaged ex post to calculate CCs. In this paper, we refer to these approaches as AVGCC, while PJM refers to a similar approach as Class AVGCC \cite{pjm_manual_20a_2025,pjm_elcc_education_2024}.
 Such approaches preserve more interannual weather variation than methods based on a single deterministic scenario or a fixed set of critical hours, but they do not fully reflect the fact that long-term investment decisions must be made before weather and load conditions are realized. As a result, the treatment of weather remains partially separated from the underlying planning decisions, rather than being integrated directly into the accreditation framework itself.

These considerations point to the need for a stochastic treatment of capacity accreditation. In long-term resource adequacy analysis, investment decisions must be made before future weather and load conditions are known, while operating decisions can adapt after uncertainty is realized. A two-stage stochastic planning model captures this structure naturally: first-stage decisions represent long-term investment choices, and second-stage decisions represent operational recourse under each realization of uncertainty. This structure explicitly enforces non-anticipativity: the same investment decision must be feasible across all scenarios, without foreknowledge of future weather and load conditions. Such models are well suited to generation expansion planning and have been used extensively in the literature \cite{peker2018two,scott2021long,9446221}. In the context of accreditation, the distinction is important. A deterministic or class average method may evaluate a fixed portfolio under multiple scenarios, but it does not optimize investment under uncertainty in the same way as a two-stage stochastic program.

A standard benchmark in stochastic programming is the \emph{expected-value} approach, under which uncertain inputs are replaced by their expected values and the resulting deterministic model is solved. We refer to the associated accreditation values as EVCC. Although computationally convenient, EVCC ignores the tails of the weather-load distribution and may therefore misrepresent adequacy-critical conditions.

\subsection{Research gap and contributions}
These limitations matter because weather affects renewable output, electricity demand, and thermal availability through temperature-dependent deratings and outages. Methods based on single historical realizations or ex post averaging may therefore fail to capture relevant weather information. This motivates a capacity accreditation framework that represents weather uncertainty directly and consistently in planning.

In this paper, we propose a stochastic capacity expansion framework for capacity accreditation that explicitly incorporates uncertainty in wind, solar, demand, and temperature-dependent thermal outages, extending the deterministic framework in \cite{9473022}, which is an economically efficient capacity accreditation framework based on the equivalent reliability enhancement capability (EREC) definition. EREC has also been implemented as a post-processing reliability method in the open-source model \cite{hope_model}.

We therefore define \emph{stochastic capacity credits} (SCC) as the credits obtained from a stochastic optimization model. SCC differs conceptually from DCC, AVGCC, and EVCC because it values resources under a non-anticipative investment decision and a scenario-dependent operational recourse model.

We further use an assessment framework to quantify the value of weather uncertainty in evaluating power system reliability by comparing SCC, DCC, AVGCC, and EVCC. 

Specifically, we answer the following research questions:

\begin{itemize}

    \item \textit{Question 1}: How can we estimate  weather uncertainty impacts on resource accreditation and resource adequacy planning (in terms of capacity values and systems generation mix)? 
    
    \item \textit{Question 2}: 
    How do different CC calculations - such as SCC, DCC,  AVGCC and EVCC - differ in their ability to capture weather-driven uncertainty and incentivize more effective resource planning?

\end{itemize}

This research contributes to the literature in several ways. First, we propose a stochastic capacity accreditation framework that quantifies how uncertainty in wind, solar, demand, and temperature-dependent thermal availability affects capacity accreditation within a stochastic programming setting. Second, we use an assessment framework, consistent with current capacity-market practice, to compare SCC against DCC, AVGCC, and EVCC under an equivalent planning reserve margin. Third, we show that SCC yields the lowest EUE among these approaches under the common benchmark, indicating improved reliability performance. We further show that incorporating weather uncertainty affects both accredited capacity values and capacity expansion decisions. In addition, we show that the equivalence between energy-only and energy-capacity market investment outcomes can be extended from the deterministic setting in \cite{9473022} to a stochastic framework.

In summary, we show that including weather uncertainty in a stochastic framework affects capacity expansion decisions and CCs, leading to more reliable resource portfolios and more informative CCs.

The remainder of this paper is organized as follows. Section~\ref{sec:method} presents the methodology, Section~\ref{sec:case study and results} describes the case study and discusses the results, and Section~\ref{sec:conclusion} concludes with key findings and directions for future research.

\section{Methodology}
\label{sec:method}
In this section, we present the stochastic capacity expansion models proposed in this study, including an energy-only market model without a price cap and an Energy–Capacity market model with a price cap and reliability constraints. We then describe and compare the different CC approaches examined in the analysis, namely, SCC, DCC, AVGCC and EVCC. Finally, we introduce the assessment framework and outline the experimental design used to evaluate the performance of SCC, DCC, AVGCC, and EVCC under the same reserve margin.

\subsection{Stochastic models of capacity expansion}
\label{subsec:stochastic models}
We implement a two-stage stochastic programming approach for long-term planning and resource adequacy problems, building upon the central capacity planning optimization model. The objective function minimizes the sum of construction costs and expected operating costs. The first stage minimizes the construction costs for conventional, wind, and solar generation, as well as storage. In the second stage, the objective is to minimize expected operating costs while accounting for uncertainty in renewable generation, temperature-dependent derating of thermal units, demand, and penalties for unserved energy. 

Regarding constraints, the model includes operational constraints for generation, storage, and load.

We model two market types:

\subsubsection{Energy-Only (EO) Market}
In this market, no price cap is imposed; instead, prices are allowed to rise to the value of lost load (VOLL) when loads cannot be served, thereby incentivizing system reliability through high prices.

\subsubsection{Energy-Capacity (EC) Market} Here, a capacity market is introduced along with a price cap in the energy market. The capacity market provides system reliability through additional capacity revenues to generators, based on their estimated CCs.

\vspace{-0.5em}

\subsection{Temperature-Derating for thermal generation}
\begin{figure}[!t]
    \centering
    \includegraphics[width=0.45\textwidth]
    {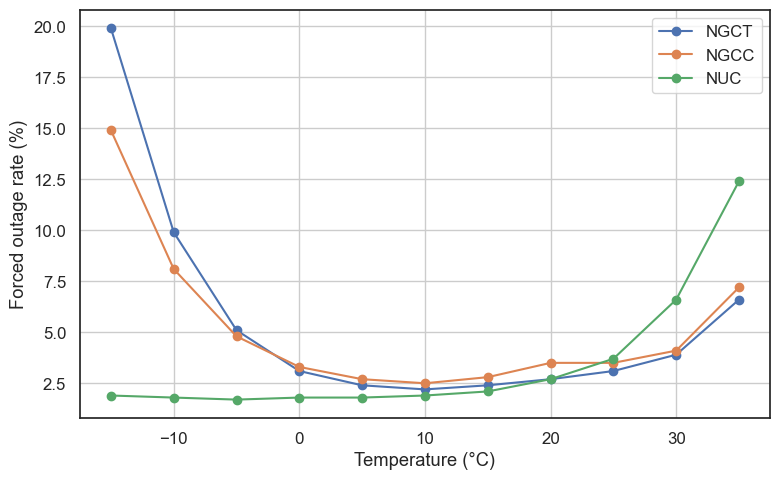}
    \caption{Temperature-dependent forced-outage rates for natural-gas
    combustion turbine (NGCT), natural-gas combined-cycle (NGCC), and
    nuclear (NUC) generation technologies.}
    \label{fig:Tem_derating}
\end{figure}

In this study, we construct a piecewise-linear function to represent the impact of temperature on thermal generation availability, based on estimated temperature–outage relationships for different thermal generation technologies from \cite{MURPHY2020114424}. These relationships are derived from  temperature correlated outages using a non-homogeneous Markov chain  \cite{MURPHY2019113513}. We refine the original discrete five-degree derating representation for each technology to a one-degree resolution by applying interpolation between consecutive five-degree data points as shown in Fig. \ref{fig:Tem_derating}. This interpolation provides a more detailed representation of temperature-dependent thermal derating while preserving the original estimates from the five-degree data.

The preceding discussion distinguishes the two modeling frameworks at a conceptual level. We now formalize these models mathematically. The objective function and constraints are presented below. 
\subsection{Objective function}

The objective is to minimize the sum of first-stage investment costs and the expected value of the second-stage operational costs over a set of scenarios $\omega \in \Omega$. The formulation is as follows:

\begin{equation}
\label{eq:main_obj}
\begin{aligned}
\min_{\substack{x_g,\,x_s}} \Bigg\{
& \sum_{g \in G^F \cup G^W \cup G^{PV}} C_g x_g
+ \sum_{s \in S} C_s x_s \\
& + \sum_{\omega \in \Omega} \pi_\omega Q_\omega(x)
\Bigg\}.
\end{aligned}
\end{equation}

The second-stage cost function $Q_\omega(\cdot)$ for each scenario $\omega$, given the first-stage decisions, is:
\begin{equation}
\label{eq:second_stage_obj}
\begin{split}
Q_\omega(x) =
\min_{\substack{
p_{g,t,\omega},\, d_{s,t,\omega},\, c_{s,t,\omega},\\
ue_{t,\omega},\, e_{s,t,\omega}
}}
\Bigg\{ 
& \sum_{\substack{g \in G,\\ t \in T}} VC_g \cdot p_{g,t,\omega} \\
& + \sum_{\substack{s \in S,\\ t \in T}} 
VSC_s \cdot \left(d_{s,t,\omega} + c_{s,t,\omega}\right) \\
& + \sum_{t \in T} VOLL \cdot ue_{t,\omega}
\Bigg\}
\end{split}
\end{equation}

subject to constraints \eqref{cons:energy_balance}--\eqref{cons:storage_energy_transition}.

\subsection{Constraints}

For conciseness, we present the constraints for the stochastic optimization model; the corresponding deterministic formulation is obtained by dropping the scenario index $\omega \in \Omega$. The model is subject to the following operational and planning constraints for all time periods $t \in T$ and scenarios $\omega \in \Omega$.

\subsubsection{Energy Balance}
\begin{equation}
\sum_{g \in G} p_{g,t,\omega} 
+ \sum_{s \in S}\left(d_{s,t,\omega} - c_{s,t,\omega}\right) 
+ ue_{t,\omega} 
= D_{t,\omega}
\label{cons:energy_balance}
\end{equation}
$\forall t \in T, \forall \omega \in \Omega$.

\subsubsection{Generation Capacity Limits}
\begin{equation}
p_{g,t,\omega} \leq x_g
\label{cons:conventional_capacity}
\end{equation}
$\forall g \in G^F, \forall t \in T, \forall \omega \in \Omega$.

\begin{equation}
p_{g,t,\omega} \leq x_g \cdot \phi_{g,t,\omega}
\label{cons:renewable_capacity}
\end{equation}
$\forall g \in G^W \cup G^{PV}, \forall t \in T, \forall \omega \in \Omega$.

\subsubsection{Thermal Derating Constraint}
\begin{equation}
p_{g,t,\omega} \leq x_g \cdot \left(1 - TDF_{g,t,\omega}\right)
\label{cons:conventional_reserve_limit}
\end{equation}
$\forall g \in G^F, \forall t \in T, \forall \omega \in \Omega$.

\subsubsection{Storage Constraints}

\paragraph{Initial and Terminal State-of-Charge Conditions}
\begin{equation}
e_{s,0,\omega} = \alpha_s \cdot x_s
\label{cons:storage_initial_soc}
\end{equation}

\begin{equation}
e_{s,|T|,\omega} = \alpha_s \cdot x_s
\label{cons:storage_terminal_soc}
\end{equation}
$\forall s \in S, \forall \omega \in \Omega$.

\paragraph{Charging and Discharging Power Limit}
\begin{equation}
\frac{c_{s,t,\omega}}{SC_s} \leq x_s
\label{cons:storage_charge_limit}
\end{equation}

\begin{equation}
\frac{d_{s,t,\omega}}{SD_s} \leq x_s
\label{cons:storage_discharge_limit}
\end{equation}
$\forall s \in S, \forall t \in T, \forall \omega \in \Omega$.

\paragraph{Storage Energy Capacity Limit}
\begin{equation}
e_{s,t,\omega} \leq HD \cdot x_s
\label{cons:storage_energy_capacity_limit}
\end{equation}
$\forall s \in S, \forall t \in T, \forall \omega \in \Omega$.

\paragraph{Storage Energy Transition}
\begin{equation}
e_{s,t,\omega} 
= e_{s,t-1,\omega} 
+ \eta_c \cdot c_{s,t,\omega} 
- \frac{1}{\eta_d} \cdot d_{s,t,\omega}
\label{cons:storage_energy_transition}
\end{equation}
$\forall s \in S, \forall t \in T \setminus \{0\}, \forall \omega \in \Omega$.

Constraint~\eqref{cons:energy_balance} represents the system-wide energy balance condition. Constraints~\eqref{cons:conventional_capacity}--\eqref{cons:renewable_capacity} impose generation capacity limits for conventional and renewable resources. Constraint~\eqref{cons:conventional_reserve_limit} further restricts thermal output through temperature-dependent derating. The storage constraints, \eqref{cons:storage_initial_soc}--\eqref{cons:storage_energy_transition}, specify the initial and terminal state of charge, charging and discharging power limits, energy capacity bounds, and intertemporal state-of-charge evolution.

\subsubsection{Energy-Capacity Market Adequacy Constraints}

The model described in the previous section, representing the energy-only market, assumes a VOLL of \$15{,}000/\text{MWh} and does not include explicit resource adequacy constraints. 
In contrast, we extend the framework by adding constraint~\eqref{cons:ec_market_adequacy_consts}, 
which ensures system reliability. At the same time, we reduce the price cap from the assumed VOLL of \$15{,}000/\text{MWh} to \$5{,}000/\text{MWh}. This constraint requires the total accredited capacity from generation resources (G) and storage units (S), calculated using their respective CCs, to meet the planning peak demand (PD) plus the planning reserve margin (PRM).

\begin{IEEEeqnarray}{rCl}
\sum_{g \in G^F \cup G^W \cup G^{PV}} \left(x_g\cdot CC_g\right) 
\nonumber\\
 + \sum_{s \in S} \left(x_s\cdot CC_s\right)
& \geq &
PD \cdot (1 + PRM)
\label{cons:ec_market_adequacy_consts}
\end{IEEEeqnarray}

\subsection{Capacity accreditation methods under stochastic and deterministic models}

We evaluate four capacity-credit methodologies within two capacity expansion frameworks: a two-stage stochastic model and a deterministic model. The stochastic model is used to compute SCC, while the deterministic model is used to compute DCC, AVGCC, and EVCC under different treatments of uncertainty.

Following the CC definition in \cite{9473022}, we first solve the relevant stochastic or deterministic capacity expansion model to obtain the optimal installed capacities and the corresponding baseline unserved energy, denoted by $EUE^{0}$. We then perturb the system by adding or subtracting 1~MW of resource $g$ and recompute the resulting unserved energy, denoted by $EUE^{g}$. Finally, we add or subtract 1~MW of perfectly reliable capacity to the optimal resource mix and compute the resulting unserved energy, denoted by $EUE^{pc}$. The generic capacity-credit definition is
\begin{equation}
CC_g = \frac{EUE^{0} - EUE^{g}}{EUE^{0} - EUE^{pc}}
\label{equa:cc_equation}
\end{equation}

The different accreditation methods are defined as follows. Each method computes CCs using a distinct optimization model.

\begin{itemize}

    \item \textbf{SCC:}  
    SCC is computed using the two-stage stochastic optimization model described in Section~\ref{subsec:stochastic models}. Only one set of first-stage investment decisions is allowed, and it must adapt across all future scenarios. The stochastic capacity credit of resource $g$ is 
    \begin{equation}
    SCC_g =
    \frac{
    \sum_{\omega \in \Omega}\sum_{t \in T}\pi_{\omega} ue^{0}_{t,\omega}
    -
    \sum_{\omega \in \Omega}\sum_{t \in T}\pi_{\omega} ue^{g}_{t,\omega}
    }{
    \sum_{\omega \in \Omega}\sum_{t \in T}\pi_{\omega} ue^{0}_{t,\omega}
    -
    \sum_{\omega \in \Omega}\sum_{t \in T}\pi_{\omega} ue^{pc}_{t,\omega}
    }.
    \end{equation}
    Here, $EUE^0$, $EUE^g$, and $EUE^{pc}$ are expected unserved energy values across all scenarios under the baseline, perturbed-resource, and perfectly reliable benchmark cases, respectively.

    \item \textbf{DCC:}  
    The deterministic capacity credit uses the same formulation as \eqref{equa:cc_equation}, but the model is solved as a deterministic optimization problem for one historical scenario year. This approach assumes perfect foresight with respect to that single realization and ignores the other weather outcomes that may occur in the future. 

    \item \textbf{AVGCC:}  
    In this approach, each uncertainty scenario is treated as an independent deterministic case, and a separate deterministic optimization model is solved for each scenario. The average capacity credit is then computed as
    \begin{equation}
    AVGCC_g
    =
    \frac{
    \frac{1}{|\Omega|}\sum_{i=1}^{|\Omega|} \left(EUE_i^{0}-EUE_i^{g}\right)
    }{
    \frac{1}{|\Omega|}\sum_{i=1}^{|\Omega|} \left(EUE_i^{0}-EUE_i^{pc}\right)
    }
    \end{equation}
    Thus, AVGCC measures the average reduction in unserved energy across scenarios relative to the average contribution of a perfectly reliable benchmark resource.

    \item \textbf{EVCC:}  
    EVCC uses the same formulation as \eqref{equa:cc_equation}, but replaces uncertain inputs with their expected values. In particular, demand, thermal derating, wind availability, and solar availability are fixed at their expected values across all scenarios and then used in a single deterministic optimization model.

\end{itemize}

We summarize the accreditation methodologies and their corresponding ISO and research applications in Table~\ref{tab:method_summary_application}.

\begin{table*}[!t]
\centering
\captionsetup{
    justification=centering,
    singlelinecheck=false,
    labelsep=newline
}
\caption{\textsc{Summary of Accreditation Methods and Applications}}
\label{tab:method_summary_application}
\footnotesize
\renewcommand{\arraystretch}{1.10}
\setlength{\tabcolsep}{4pt}

\resizebox{\textwidth}{!}{
\begin{tabular}{l c c c p{9.8cm}}
\hline
Method 
& Data Input 
& Aleatoric Uncertainty 
& Expansion Model Type 
& \multicolumn{1}{c}{Research and Application} \\
\hline

SCC 
& Multiple scenarios 
& Yes 
& Stochastic 
& Proposed in this research. \\

DCC
& Single historical scenario 
& No 
& Deterministic 
& MISO (Thermal)~\cite{miso_resource_accreditation_v21_2024}, 
\cite{9473022}, \cite{9729557}, \cite{frew20178760based}, 
\cite{PECORA2025138979}. \\

AVGCC 
& Individual historical scenarios 
& Partial 
& Deterministic 
& PJM (Thermal, VRE)~\cite{pjm_manual_20a_2025,pjm_elcc_education_2024}; 
NYISO (Thermal, VRE)~\cite{nyiso_icap_manual_2026}; 
SPP (VRE)~\cite{spp2024elcc,spp_elcc_study_scope_2025}; 
NESO (VRE)~\cite{emr_ecr_report}; 
MISO~\cite{miso_resource_accreditation_v21_2024}; 
\cite{AWARA2023100021,li2026riskbasedcapacityaccreditationresourcecolocated,LI2026112165}. \\

EVCC 
& Expected-value inputs 
& No 
& Deterministic 
& \cite{9729557}, \cite{spp_sawg_wind_solar_2017}. \\

\hline
\end{tabular}
}
\end{table*}

\subsection{Assessment framework}
\label{subsubsec:assessment framework}
\begin{figure}[htb!]
    \centering
    \includegraphics[width=0.5\textwidth]{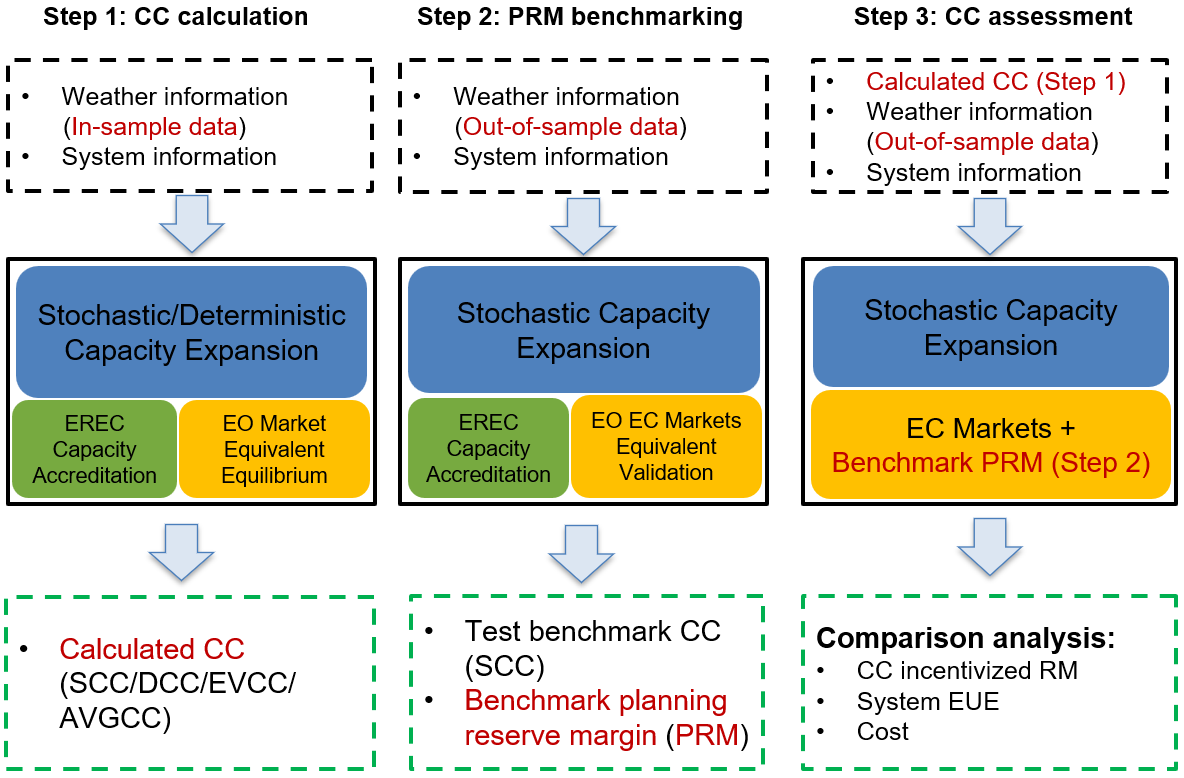}
    \caption{Assessment framework}
    \label{fig:assessment_framework}
\end{figure}

The assessment framework is described below. We use a three-step assessment framework to compare alternative capacity accreditation methods on a common, economically meaningful basis that is consistent with current market practice as shown in Fig. \ref{fig:assessment_framework}.

\textbf{Step 1: CC calculation.} We use the in-sample dataset to calculate various CCs from the EO model. These calculated capacity credits are then carried forward into the out-of-sample evaluation in step 3.

\textbf{Step 2: PRM benchmarking.} We derive a benchmark planning reserve margin (PRM) from the out-of-sample benchmark resource mix (RM) and its associated benchmark capacity credits. PRM is used because it is consistent with current market practice and provides a transparent common adequacy target for comparing accreditation methods \cite{pjm_manual_20a_2025}. This benchmark PRM is then imposed in the adequacy constraint of the evaluation model. By fixing the same PRM across all cases, differences in reliability and cost can be attributed to the accreditation method itself, rather than to differences in the adequacy standard.

\textbf{Step 3: CC assessment.} We perform an out-of-sample assessment using an independent set of weather and demand scenarios. This out-of-sample evaluation is necessary because assessing all methods on the same data used to construct the accreditation values may bias the comparison in favor of methods that are more closely fitted to the in-sample realizations. Accordingly, the out-of-sample scenarios and the benchmark PRM are held fixed across all cases, while the accreditation method varies across SCC, DCC, AVGCC, and EVCC. Each method is then evaluated within the same stochastic energy--capacity market framework, allowing a consistent comparison of resource mix, expected unserved energy and system cost.

\begin{table}[!t]
\centering
\captionsetup{
    justification=centering,
    singlelinecheck=false,
    labelsep=newline
}
\caption{\textsc{Technical--Economic Parameters of the Considered Technologies}}
\label{tab:techno_economic_params}
\footnotesize
\renewcommand{\arraystretch}{1.08}
\setlength{\tabcolsep}{6pt}

\begin{tabular}{lccc}
\hline
Technology
& Investment Cost
& Variable Cost
& FOR \\
& (\$/MW-yr)
& (\$/MWh)
& (\%) \\
\hline

NGCT          
& 104,472 & 30.5 & 2.80 \\

NGCC          
& 113,641 & 21.6 & 3.30 \\

Nuclear       
& 399,322 & 10.5 & 1.60 \\

Wind (1--3)   
& 86,843  & --   & --   \\

Solar (1--3)  
& 58,221  & --   & --   \\

Storage       
& 87,624  & 0.01 & --   \\

\hline
\end{tabular}
\end{table}

\section{Case study and results}
\label{sec:case study and results}
In this section, we first describe the assumptions and case study setup. We then show simulation results to address the research questions. 

\subsection{Assumptions}
We assume price-taking behavior and a single-node system without transmission constraints. If transmission constraints were included, it would be more difficult to distinguish whether the observed effects were driven by weather uncertainty itself or by network and power-flow limitations. We also assume inelastic demand, simplified unit commitment, and continuous one-shot investment decisions. All six technologies: NGCT, NGCC, nuclear, wind, solar, and storage are available for investment, while operational availability is represented through resource-specific availability factors and temperature-dependent derating. These assumptions are appropriate for a long-term strategic planning study focused on accreditation and adequacy rather than detailed operational market behavior. 

\subsection{Case study setup}
We set up a greenfield (i.e., no existing capacity) stochastic capacity expansion model. Five representative years (2015-2019) of time-series data, including hourly profiles of load, temperature, wind, and solar, are used as the in-sample dataset, while three years of data (2021-2023) are used as the out-of-sample test set. For each year in the case study, the peak load is normalized to 10{,}000~MW, and the original hourly load profiles are rescaled accordingly. All load data are based on the ERCOT system.

For temperature data, we use location-specific temperature observations obtained from the Open-Meteo database\cite{openmeteo_website}. For VRE resources, we select three wind sites (\textit{wind1}, \textit{wind2}, and \textit{wind3}) and three solar sites (\textit{solar1}, \textit{solar2}, and \textit{solar3}) to reflect geographic diversity. Hourly wind and solar availability factors are obtained from the Renewable Ninja database \cite{renewables_ninja} based on the corresponding site locations. Across the three locations, the average availability factors for wind resources are 0.2867, 0.3668, and 0.4635. At the three respective solar locations, the values were 0.1798, 0.1832, and 0.2097.

We model six types of technology: NGCT, NGCC, nuclear, wind, solar, and storage. All resource characteristics and cost assumptions are summarized in Table~\ref{tab:techno_economic_params}. Cost parameters are derived from the U.S. Energy Information Administration (EIA) and the National Laboratory of the Rockies Annual Technology Baseline (ATB) \cite{nrel_atb_2024_technologies,eia_aeo2023_cost_perf}. The VOLL and price cap are set to \$15{,}000/MWh and \$5{,}000/MWh, respectively.

\begin{table*}[!t]
\centering
\captionsetup{
    justification=centering,
    singlelinecheck=false,
    labelsep=newline
}
\renewcommand{\arraystretch}{1.05}
\setlength{\tabcolsep}{4pt}
\footnotesize

\begin{minipage}[t]{0.49\textwidth}
\centering
\caption{\textsc{Wind Capacity Credits and Percentage Changes Relative to SCC}}
\label{tab:wind_cc_parallel}

\resizebox{\linewidth}{!}{
\begin{tabular}{lcccccc}
\hline
\multirow{2}{*}{Method} 
& \multicolumn{3}{c}{Capacity Credits} 
& \multicolumn{3}{c}{\% Changes Relative to SCC} \\
\cline{2-4}\cline{5-7}
& Wind1 & Wind2 & Wind3 & Wind1 & Wind2 & Wind3 \\
\hline

SCC
& 0.1879 & 0.1048 & 0.0240
& 0.00\% & 0.00\% & 0.00\% \\

AVGCC
& 0.2496 & 0.1076 & 0.0598
& 32.84\% & 2.67\% & 149.17\% \\

EVCC
& 0.1798 & 0.1419 & 0.1624
& -4.31\% & 35.40\% & 576.67\% \\

DCC (2019)
& 0.0380 & 0.0060 & 0.0120
& -79.78\% & -94.27\% & -50.00\% \\

DCC (2018)
& 0.1480 & 0.0030 & 0.0060
& -21.23\% & -97.14\% & -75.00\% \\

DCC (2017)
& 0.2350 & 0.0940 & 0.0105
& 25.07\% & -10.31\% & -56.25\% \\

DCC (2016)
& 0.4060 & 0.1970 & 0.1000
& 116.07\% & 87.98\% & 316.67\% \\

DCC (2015)
& 0.3260 & 0.2310 & 0.0155
& 73.50\% & 120.42\% & -35.42\% \\

\hline
\end{tabular}
}
\end{minipage}
\hfill
\begin{minipage}[t]{0.49\textwidth}
\centering
\caption{\textsc{Solar Capacity Credits and Percentage Changes Relative to SCC}}
\label{tab:solar_cc_parallel}

\resizebox{\linewidth}{!}{
\begin{tabular}{lcccccc}
\hline
\multirow{2}{*}{Method} 
& \multicolumn{3}{c}{Capacity Credits} 
& \multicolumn{3}{c}{\% Changes Relative to SCC} \\
\cline{2-4}\cline{5-7}
& Solar1 & Solar2 & Solar3 & Solar1 & Solar2 & Solar3 \\
\hline

SCC
& 0.2115 & 0.1906 & 0.1636
& 0.00\% & 0.00\% & 0.00\% \\

AVGCC
& 0.1269 & 0.1177 & 0.1370
& -40.00\% & -38.25\% & -16.26\% \\

EVCC
& 0.2188 & 0.2052 & 0.2350
& 3.45\% & 7.68\% & 43.60\% \\

DCC (2019)
& 0.0150 & 0.0290 & 0.0790
& -92.91\% & -84.78\% & -51.72\% \\

DCC (2018)
& 0.1570 & 0.1030 & 0.1530
& -25.76\% & -45.96\% & -6.50\% \\

DCC (2017)
& 0.0990 & 0.0980 & 0.1165
& -53.18\% & -48.58\% & -28.80\% \\

DCC (2016)
& 0.0200 & 0.0350 & 0.0710
& -90.54\% & -81.64\% & -56.61\% \\

DCC (2015)
& 0.0225 & 0.0375 & 0.0870
& -89.36\% & -80.33\% & -46.83\% \\

\hline
\end{tabular}
}
\end{minipage}

\end{table*}

\begin{figure*}[!t]
    \centering

    \begin{minipage}[t]{0.48\textwidth}
        \centering
        \includegraphics[width=\linewidth]
        {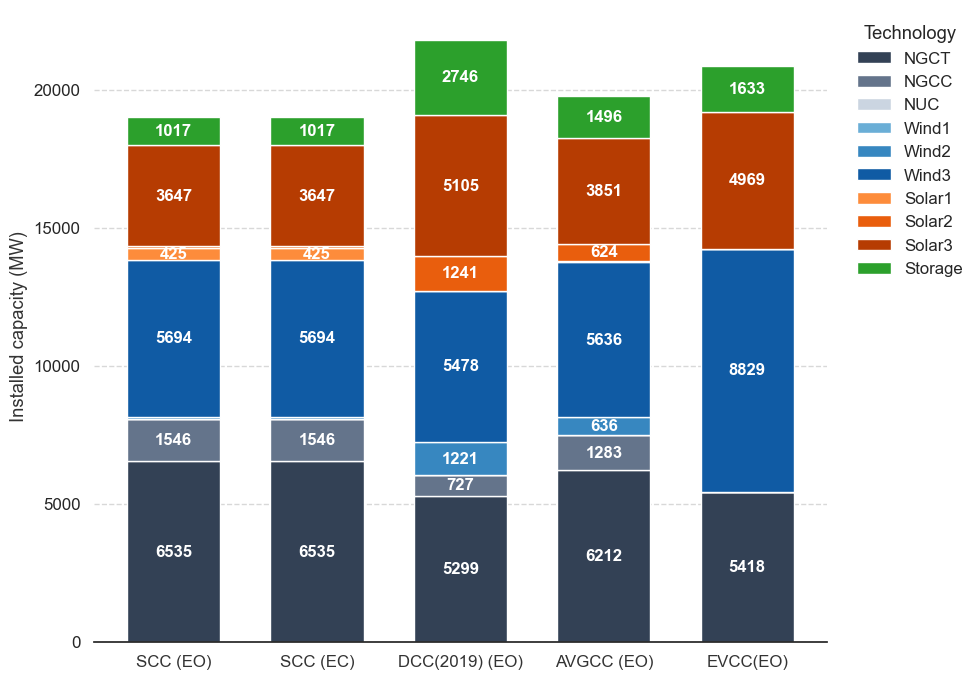}
        \caption{Installed capacity by technology under in-sample
        uncertainty with different methods.}
        \label{fig:step_1_resource}
    \end{minipage}
    \hfill
    \begin{minipage}[t]{0.48\textwidth}
        \centering
        \includegraphics[width=\linewidth]
        {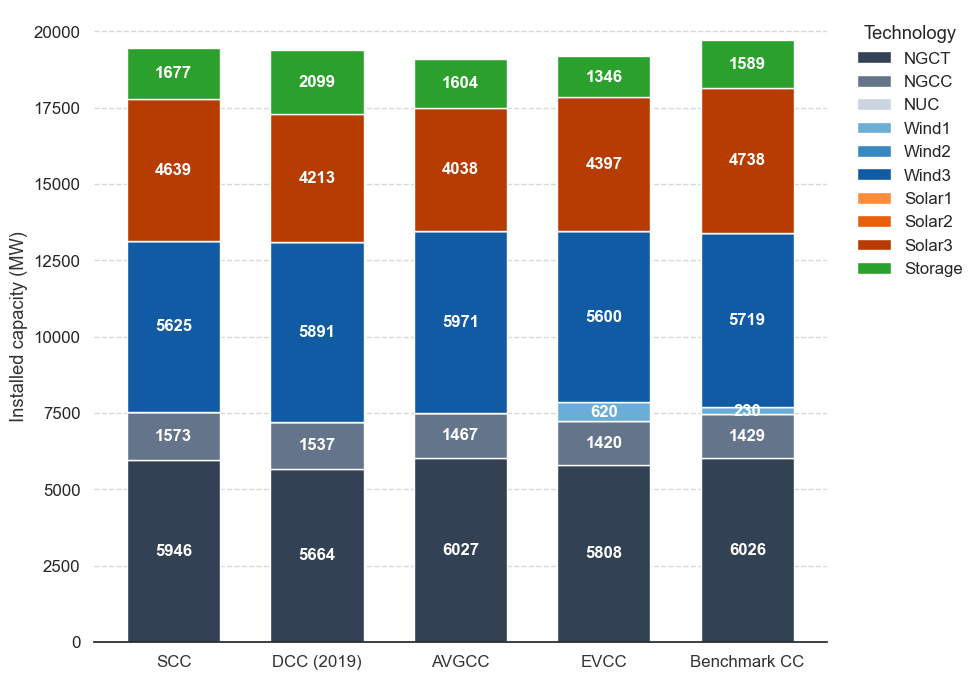}
        \caption{Installed capacity by technology under different
        capacity-credit incentives in the out-of-sample
        energy--capacity market.}
        \label{fig:incentivize_resource_mix}
    \end{minipage}

\end{figure*}

\subsection{Impacts of CC calculation approach on resource planning}

Fig.~\ref{fig:step_1_resource} shows the optimal resource mix under different modeling methods. From left to right, the bars represent the EO market equilibrium under \textit{SCC (EO)}, the SCC-incentivized EC market equilibrium under \textit{SCC (EC)}, the EO market resource planning results under \textit{DCC (2019)} as a representative Deterministic method case, the EO market resource planning results under \textit{AVGCC (EO)} methods, and the EO market planning results under \textit{EVCC}. The EO market without a price cap can generally be regarded as a benchmark for long-term market equilibrium or the central planning outcome. Here, the first two bars show identical resource mixes, extending the deterministic result in ~\cite{9473022} to a stochastic setting and demonstrating that market design equivalence continues to hold. These numerical results demonstrate that EREC-based SCC accreditation provides more accurate investment incentives while capturing weather uncertainty.

Relative to the stochastic benchmark, the \textit{DCC (2019) (EO)} method shifts the portfolio away from thermal capacity, particularly NGCT and NGCC, and toward Wind1, Solar2, Solar3, and storage, while slightly reducing Wind3. This pattern arises because investment decisions are based on a single historical weather year treated as a fully known deterministic input. This indicates that VRE generation and storage are scheduled with perfect foresight over the realized 2019 conditions. Consequently, the specific weather and load conditions in 2019 make selected VRE and storage resources appear more valuable for adequacy than in the stochastic benchmark. The \textit{AVGCC (EO)} method produces a similar but more moderate shift, especially away from NGCC and toward Wind2, Solar2, Solar3, and storage. Here, each deterministic run also assumes perfect foresight within the corresponding scenario, which tends to favor greater investment in VRE and storage. The largest distortion appears in the \textit{EVCC (EO)} method, where NGCC is eliminated entirely, NGCT declines further, and the portfolio becomes heavily concentrated in Wind3, Solar3, and storage because weather and load variability are compressed into a single expected representation that smooths scarcity-critical conditions.

Methods that incorporate multi-year weather information produce more balanced portfolios, whereas deterministic or average-based methods that ignore weather uncertainty distort investment toward selected renewable and storage resources.

\subsection{Capacity credit calculation under different accreditation methods}

Tables~\ref{tab:wind_cc_parallel} and~\ref{tab:solar_cc_parallel} report wind and solar capacity credits under alternative accreditation approaches. The upper panel entries correspond to SCC, AVGCC and EVCC, while the remaining rows report the single-year deterministic cases from DCC (2019) to DCC (2015). Across these single-year deterministic methods, VRE capacity credits vary markedly from one weather year to another. For example, wind and solar credits in several locations range from nearly zero in some years to substantially higher values in others. This indicates that accreditation based on a single historical realization can produce unstable signals for future capacity payments and long-term resource planning.

Among the alternative methods, EVCC exhibits the largest distortions relative to SCC, particularly for Wind2, Wind3, and Solar3, while also overstating the capacity value of some higher-average-output resources. This occurs because EVCC replaces weather and load variability with a single expected representation, thereby smoothing scarcity-critical hours and obscuring the joint conditions that determine adequacy value. AVGCC is more comprehensive than the single-year deterministic cases because it incorporates multiple weather realizations, but it still averages reliability contributions ex post and therefore masks resource-specific and interannual variation. By contrast, the SCC results suggest that preserving the full multi-year weather chronology yields a more stable estimate of CC value than either expected-value or single-year accreditation.

For thermal generation, Table~\ref{tab:thermal_cc_temp_derating} reports the effects of temperature-dependent derating under alternative accreditation approaches. From top to bottom, the rows correspond to the different accreditation methods; in addition, the second row presents the SCC method without temperature-dependent derating. In that specification, thermal availability is represented using a fixed forced outage rate based on the unconditional outage rates reported in \cite{MURPHY2020114424}. Comparing this method with the SCC method that includes temperature-dependent derating shows that neglecting derating systematically overstates thermal capacity credits. The expectation is that thermal resources provide their greatest reliable capacity during periods of extreme heat or cold, however, the temperature effects reduce their effective available capacity. Nevertheless, despite these numerical differences, thermal capacity credits remain relatively similar across the different accreditation approaches.

\begin{table}[!t]
\centering
\captionsetup{
    justification=centering,
    singlelinecheck=false,
    labelsep=newline
}
\caption{\textsc{Thermal Capacity Credits and Percentage Changes Relative to SCC}}
\label{tab:thermal_cc_temp_derating}
\footnotesize
\renewcommand{\arraystretch}{1.08}
\setlength{\tabcolsep}{3.5pt}

\resizebox{\columnwidth}{!}{
\begin{tabular}{lcccccc}
\hline
\multirow{2}{*}{Method}
& \multicolumn{3}{c}{Capacity Credits}
& \multicolumn{3}{c}{\% Changes Relative to SCC} \\
\cline{2-4}\cline{5-7}
& NGCT 
& NGCC 
& NUC 
& NGCT 
& NGCC 
& NUC \\
\hline

SCC
& 0.9446 & 0.9401 & 0.8993
& 0.00\% & 0.00\% & 0.00\% \\

SCC (No TDF)
& 0.9720 & 0.9670 & 0.9840
& 2.90\% & 2.86\% & 9.42\% \\

AVGCC
& 0.9421 & 0.9373 & 0.8934
& -0.26\% & -0.30\% & -0.66\% \\

EVCC
& 0.9436 & 0.9387 & 0.8973
& -0.10\% & -0.15\% & -0.22\% \\

DCC (2019)
& 0.9340 & 0.9280 & 0.8760
& -1.12\% & -1.29\% & -2.59\% \\

DCC (2018)
& 0.9390 & 0.9340 & 0.8880
& -0.59\% & -0.65\% & -1.26\% \\

DCC (2017)
& 0.9475 & 0.9435 & 0.9050
& 0.31\% & 0.36\% & 0.63\% \\

DCC (2016)
& 0.9500 & 0.9470 & 0.9110
& 0.57\% & 0.73\% & 1.30\% \\

DCC (2015)
& 0.9395 & 0.9340 & 0.8875
& -0.54\% & -0.65\% & -1.31\% \\

\hline
\end{tabular}
}
\end{table}

\subsection {Performance assessment of capacity credit methods}

\begin{table*}[htbp]
\centering
\captionsetup{
    justification=centering,
    singlelinecheck=false,
    labelsep=newline
}
\caption{\textsc{Out-of-Sample Performance Under the Benchmark Planning Reserve Margin}}
\label{tab:step3_results_reduced}

\setlength{\tabcolsep}{6pt}
\footnotesize
\resizebox{0.95\textwidth}{!}{
\begin{tabular}{l c c c c c c}
\toprule
 & EUE (MWh) 
 & NEUE (ppm) 
 & Cost (\$ M) (PCAP=5000) 
 & Social Cost (\$ M) (VOLL=15000) 
 & EUE Change (\%)  
 & Social Cost Change (\%) \\
\midrule
SCC
& 509.46  & 9.69   & 2362.15 & 2367.25 & -45\% & 0.11\% \\

DCC (2019)
& 1829.12 & 34.80  & 2362.29 & 2380.58 & 99\%  & 0.67\% \\

AVGCC
& 1651.43 & 31.42  & 2352.46 & 2368.98 & 79\%  & 0.18\% \\

EVCC
& 6007.75 & 114.32 & 2338.56 & 2398.64 & 552\% & 1.43\% \\

Benchmark CC
& 921.21  & 17.53  & 2355.55 & 2364.76 & 0\%   & 0.00\% \\
\bottomrule
\end{tabular}
}

\vspace{2pt}
\begin{minipage}{0.95\textwidth}
\footnotesize
\textit{Note:} The benchmark planning reserve margin is fixed at $PRM=-0.078117$ for all methods.
\end{minipage}

\end{table*}

We use the assessment framework introduced in Section~\ref{subsubsec:assessment framework} to derive the benchmark installed capacities and CCs from the out-of-sample dataset, and we calculate a planning reserve margin of $\mathrm{PRM}=-0.078117$. The planning reserve margin is derived from the out-of-sample benchmark case, where the resulting CCs and resource mix are used to estimate PRM. We only treat it as a benchmark for controlled comparison, not as an empirical recommendation for the reserve margin. The negative PRM achieved here may be due to perfect foresight of the realized load and renewable profiles for each scenario and allowance for system-level unserved energy to reflect the economic efficiency of system planning and market designs. This common benchmark is subsequently used to evaluate all accreditation methods against a consistent adequacy target.

Table~\ref{tab:step3_results_reduced} shows that the choice of accreditation method materially affects out-of-sample performance. SCC yields the strongest adequacy outcome, with EUE equal to 509.46~MWh. By contrast, DCC~(2019) performs substantially worse, with EUE equal to 1829.12~MWh, more than three times the SCC value. Although DCC~(2019) is based on an actual historical weather year, its accreditation values are still derived from a single deterministic realization and therefore remain highly sensitive to the specific weather and load conditions observed in that year. As a result, it provides a less reliable adequacy signal than SCC and leads to a higher social cost. EVCC produces the most severe under-reliability, with EUE of 6007.75~MWh, approximately 11.8 times the SCC value. This reliability issue arises because EVCC replaces the full weather--load distribution with expected-value inputs, thereby obscuring scarcity-critical conditions and overstating effective firm capacity. AVGCC performs better than EVCC but remains substantially less reliable than SCC, with EUE equal to 1651.43~MWh. Although it retains more historical information than EVCC, averaging scenario-specific contributions still masks resource-specific and scarcity-hour variation.

The cost comparison further supports this conclusion. Although SCC results in a slightly higher capped system cost, it yields the lowest social cost once unserved energy is valued at VOLL. Overall, SCC provides the most desirable balance between economic efficiency and resource adequacy because it more accurately reflects weather-driven uncertainty in capacity valuation.

Fig.~\ref{fig:incentivize_resource_mix} shows the installed resource mix under different capacity-credit inputs in the out-of-sample energy--capacity market. The results indicate that different accreditation methods provide different investment signals and therefore lead to different equilibrium resource mixes. 

Relative to SCC, DCC~(2019) induces a moderate reallocation away from thermal capacity, particularly NGCT and NGCC, and toward Wind3 and storage, while slightly reducing Solar3. This pattern is consistent with the use of a single deterministic weather year, which makes the accredited contribution of selected resources highly dependent on the realized 2019 conditions. Compared with SCC, AVGCC induces moderate shifts in the optimal expansion outcome. EVCC yields the largest distortion, characterized by lower thermal investment and substantial reallocation toward selected renewable and storage technologies. These resource mix results occur because EVCC and AVGCC do not capture weather uncertainty as effectively as SCC and therefore tend to overstate the capacity value of some VRE resources, leading to distorted planning signals under weather uncertainty. 
Overall, Fig.~\ref{fig:incentivize_resource_mix} exhibits less resource mix variation than Fig.~\ref{fig:step_1_resource}. This is because Fig.~\ref{fig:step_1_resource} compares cases with different data inputs and different optimization models, whereas Fig.~\ref{fig:incentivize_resource_mix} varies only the capacity-credit inputs while utilizing the same out-of-sample dataset and optimization model.

\section{Conclusion}
\label{sec:conclusion}
This paper investigates how weather-driven uncertainty in wind, solar, demand, and ambient temperature affects capacity accreditation and long-term resource adequacy planning. We argue that capacity accreditation should be evaluated not only as a reliability metric, but also as an investment-incentive mechanism that shapes long-term resource mix outcomes. We develop a two-stage stochastic optimization framework that explicitly captures renewable variability, load uncertainty, and temperature-dependent thermal derating. Furthermore, we use an assessment framework to compare SCC, DCC, AVGCC, and EVCC in a price-capped energy--capacity market under the same reserve margin.

Three main findings emerge. First, weather uncertainty materially changes capacity value because wind availability, solar availability, demand, and temperature-driven thermal derating jointly determine scarcity conditions. VRE capacity values are highly sensitive to the realized weather year, showing that DCC with a single weather year can provide unstable and misleading signals. Second, incorporating temperature-dependent derating reduces the accredited capacity of thermal generators, indicating that conventional units should not be treated as perfectly firm under extreme temperature conditions and underscoring the importance of explicitly modeling thermal derating in capacity accreditation. Third, SCC provides the most informative representation of adequacy value because it preserves uncertainty in weather, load, renewable availability, and thermal derating across scenarios. By contrast, EVCC and AVGCC reduce some volatility, but their expected-value or averaging treatments can still misrepresent marginal reliability contributions during scarcity-relevant conditions. We also show that, when CCs are correctly specified, the stochastic energy--capacity market yields nearly the same optimal resource mix as the energy-only benchmark, extending the deterministic result to a stochastic setting.

The out-of-sample assessment confirms the practical value of SCC. Under the benchmark reserve margin, SCC yields substantially better reliability performance than the alternative methods. EVCC produces severe under-reliability, while AVGCC remains intermediate. Although SCC results in a slightly higher capped market cost, this increase is small and is offset by improved reliability performance. Once unserved energy is valued at VOLL, SCC achieves the lowest social cost.

Several limitations remain. The analysis relies on a limited sample of historical weather years and does not explicitly model longer-term structural uncertainties, including climate-change impacts, policy evolution, fuel-price uncertainty, and load growth. Future work will extend the framework to multi-stage planning, transmission representations, brownfield capacity accreditation settings that are more closely aligned with current capacity-market practice, and broader scenario ensembles.

\section*{Acknowledgment}
This project is supported by the MITEI Future Energy Systems Center Award. The authors thank Prof. Benjamin F. Hobbs for his continued support, suggestions and discussions. The authors also thank the MIT Writing and Communication Center for its writing guidance and advice.

\bibliographystyle{IEEEtran}
\bibliography{bi2}

\end{document}